\documentclass[doublecol]{epl2}

\usepackage{cite}
\usepackage{xcolor}
\usepackage{amssymb}
\usepackage{slashed}
\usepackage{lineno,hyperref,geometry,amsmath,appendix}
\modulolinenumbers[5]
\usepackage{bbold}
\newcommand{\gdualn}[1]{\overset{\:{}^{{}^{\boldsymbol{\neg}}}}{\smash[t]{#1}}} 
\def\O{\mbox{\boldmath$\displaystyle\mathbb{O}$}}

\def\R{\mbox{\boldmath$\displaystyle\mathbb{R}$}}

\def\x{\mbox{\boldmath$\displaystyle\boldsymbol{x}$}}
\def\y{\mbox{\boldmath$\displaystyle\boldsymbol{y}$}}

\def\I{\openone}

\def\p{\mbox{\boldmath$\displaystyle\boldsymbol{p}$}}
\def\a{\mbox{$\displaystyle\mathfrak{a}$}}
\def\da{\mbox{$\gdualn{\displaystyle\mathfrak{a}}$}}
\def\0{\mbox{\boldmath$\displaystyle\boldsymbol{0}$}}

\def\openone{\mathbb I}

\title{Spin-half bosons with mass dimension three half: Evading the spin-statistics theorem}
\shorttitle{Spin-half bosons with mass dimension three half: evading the spin-statistics theorem} 

\author{Dharam Vir Ahluwalia\inst{1} \and Cheng-Yang Lee\inst{2} }

\institute{                    
  \inst{1} Center for the Studies of the Glass Bead Game, Notting Hill, Victoria 3168, Australia\\
  \inst{2} Center for Theoretical Physics, College of Physics,
Sichuan University, Chengdu, 610064, China
}

\abstract{By exploiting the freedom in defining the dual of spinors, we report an unexpected theoretical discovery of a 
quantum field theory of spin-half bosons.
It fulfils Dirac's 1969-70 observation that ``there must be boson variables connected with electrons.'' The theory is local, Lorentz-invariant,  and has a positive-definite Hamiltonian. 
We formulate the unitarity-preserving scattering theory to accommodate the new dual and the associated adjoint.  A model of Yukawa interaction with spin-half bosons and fermions of equal masses is studied to explicitly show unitarity. }

\begin{document}

\maketitle

\begin{itemize}

\item[\empty]\textit{There must be such boson variables connected with electrons.}
\item[\empty]
\hfill {\small P. A. M. Dirac~\cite{Dirac:1970}}
\end{itemize}

\section{Introduction} 

With ninety five percent of the existence dark, in one way or another, its particle landscape  cannot be a mere copy of one sort or the other of the Standard Model of the high energy physics (SM). This may appear as an audacious thought to many physicists. But the unexpected theoretical developments since 2004~\cite{Ahluwalia:2004ab,ahluwalia_2019,AHLUWALIA20221}, coupled with a lack of empirical signature for supersymmetry, suggests to consider the unexpected as a first-principle scenario for their darkness~\cite{AHLUWALIA20221}.  

Recently, a new class of fundamental particles 
have been reported 
in~\cite{Ahluwalia:2020miz,Ahluwalia:2020jkw}.  While it is not widely known, Dirac himself argued for electrons to have a bosonic 
partner~\cite{Dirac:1970}.  Here, by exploiting  a freedom in defining the dual of spinors (and the associated adjoint), we 
 construct a theory of spin-half bosonic particles. These bosons cannot enter the SM doublets because of the statistics mismatch. The natural place for them is the dark-matter sector. As a by-product, they cancel the fermionic contributions to the cosmological constant if for every Dirac fermion of the SM there exists a here-reported bosonic partner of the same mass.
 
\section{Quantum field theory of spin-half bosons}

In the SM, the Lagrangian density, at the level of kinematics, is assumed rather than derived. The dynamics enters through the principle of local gauge invariance -- the choice of symmetry group is then chosen on the basis  of the hints hiding in the phenomenology.  
 
 One exception to this general wisdom is to derive, rather than assume, both the quantum field and the Lagrangian density. This approach was initiated in 1964 by Steven Weinberg, and reached the textbook level in his monograph -- \textit{The Quantum Theory of Fields}~\cite{Weinberg:1995mt}. 
 
 His approach put the SM on a firm theoretical footing by unifying quantum mechanics with Poincar\'e space-time symmetries. An important departure from this approach occurred  when mass dimension one fermions were first constructed~\cite{Ahluwalia:2004sz,Ahluwalia:2004ab}. Here we use the results from both approaches to construct a bosonic quantum field of spin-half, and the associated  Lagrangian density. 

 Without assuming a Lagrangian density, we begin with the rest spinors  for spin-half \textit{derived} by Weinberg in the $\mathcal{R}\oplus\mathcal{L}\vert_{j=1/2}$ representation space~\cite{Weinberg:1995mt}
\begin{equation}
\xi_1(\0)=\sqrt{m} \left[
\begin{array} {c}
1\\
0\\
1\\
0
\end{array}
\right],~
\xi_2(\0)=\sqrt{m} \left[
\begin{array} {c}
0\\
1\\
0\\
1
\end{array}
\right], \label{eq:xi12}
\end{equation} 
and
\small{
\begin{equation}
\xi_3(\0)=\sqrt{m} \left[
\begin{array} {c}
0\\
1\\
0\\
-1
\end{array}
\right],~
\xi_4(\0)=\sqrt{m} \left[
\begin{array} {c}
-1\\
0\\
1\\
0
\end{array}
\right].\label{eq:xi34}
\end{equation} 
 These coincide with the eigenspinors of 
$\Omega_5=\gamma_0$, the fifth square root of the identity matrix in the $\mathcal{R}\oplus\mathcal{L}\vert_{j=1/2}$ representation space~\cite{Ahluwalia:2020jkw}. The  $\xi_i(\p)$ for an arbitrary momentum are obtained by acting the
$\mathcal{R}\oplus\mathcal{L}\vert_{j=1/2}$ boost operator 
\begin{equation}
\mathcal{D}(\p)  = \sqrt{\frac{E+m}{2 m}}
\left[ 
\begin{array}{cc}
\I + \frac{\boldsymbol{\sigma\cdot p}}{E+m} & \O \\ 
\O & \I-\frac{\boldsymbol{\sigma\cdot p}}{E+m} 
\end{array}
\right]
\end{equation}
on $\xi_i(\0)$: 
\begin{equation}
\xi_i(\p) = \mathcal{D}(\p) \xi_i(\0), \quad i=1,\cdots,4.
\end{equation}
The crucial point of departure enters by defining a new dual ~\cite{Ahluwalia:2020jkw,HoffdaSilva:2022ixq}
\begin{align}
 \gdualn{\xi}_i(\p) & = \big[\mathcal{P} \, \xi_i(\p)\big]^\dagger 
 \gamma_0 \label{eq:nd}\\ 
& = \begin{cases}
+ \big[\xi_i(\p)\big] ^\dagger \gamma_0,\quad i=1,2\\
- \big[\xi_i(\p)\big]^\dagger \gamma_0, \quad i=3,4 
\end{cases}
\label{eq:dual_spinor}
\end{align}
where the covariant parity operator is defined as~\cite{Speranca:2013hqa,ahluwalia_2019}
\begin{equation}
\mathcal{P} \stackrel{\mathrm{def}}{=} m^{-1} \gamma_\mu p^\mu.
\end{equation}
With the dual thus defined, the orthonormality relations and spin sums are found to be
\begin{equation}
\gdualn{\xi}_i(\p) \,\xi_j(\p) = 2m \delta_{ij},
\end{equation}
and
\begin{align}
&\sum_{i=1,2} \xi_i(\p)\gdualn{\xi}_i(\p) = \gamma_\mu p^\mu  + m \I, \label{eq:12}\\ 
&\sum_{i=3,4} \xi_i(\p)\gdualn{\xi}_i(\p) = -( \gamma_\mu p^\mu  - m \I).
\label{eq:34}
\end{align}

We now use the $\xi_i(\p)$ as expansion coefficients for a spin-half quantum field
\begin{align}
\mathfrak{a}(x)\stackrel{\mathrm{def}}{=}\int&\frac{d^3 p}{(2\pi)^{3/2}}
\frac{1}{\sqrt{2E}}{\Bigg[}\sum_{i=1,2} 
a_i(\p) \xi_i(\p) e^{-i p\cdot x}\nonumber\\
&+\sum_{i=3,4} 
b^\dagger_i(\p) \xi_i(\p) e^{i p\cdot x}{\Bigg]}.
\end{align}
This field is same as the usual spin-half field, of Dirac. 
The associated statistics follows  \textit{only}  after its adjoint is defined.
Instead of the Dirac adjoint, we introduce a new adjoint that exploits 
the new duals (\ref{eq:dual_spinor})
\begin{align}
\gdualn{\mathfrak{a}}(x)\stackrel{\mathrm{def}}{=}\int&\frac{d^3 p}{(2\pi)^{3/2}}
\frac{1}{\sqrt{2 E}}{\Bigg[}\sum_{i=1,2} 
a^\dagger_i(\p) \gdualn{\xi}_i(\p) e^{i p\cdot x}\nonumber\\
&+\sum_{i=3,4} 
b_i(\p) \gdualn{\xi}_i(\p) e^{-i p\cdot x}{\Bigg]} \label{eq:na}
\end{align}
where
\begin{align}
&[a_i(\p),a_j^\dagger(\p^\prime)]_{\pm} =\delta^3\left(\p-\p^\prime\right)
\delta_{ij}, \nonumber\\
&[a_i(\p),a_j(\p^\prime)]_{\pm}=[a^\dagger_i(\p),a^\dagger_j(\p^\prime)]_{\pm}=0.
\label{eq:anti}
\end{align}
The upper and lower signs represent anti-commutator and commutator respectively.
We assume the same for $b_i(\p)$ and $b_i^\dagger(\p)$.

To establish the statistics we follow the procedure outlined in~\cite{Ahluwalia2014}. Consider two space-like separated events $x$ and $y$, and evaluate the amplitude for the particle, described by 
the $\mathfrak{a}(x)$-$\gdualn{\mathfrak{a}}(x)$ system, to propagate from $x$ to $y$ (with $y_0 >x_0$). Since we are considering space-like separations, there exist a class of observers for whom $x_0>y_0$. We call the former observers to be 
$\mathcal{O}$ and the latter to be $\mathcal{O}^\prime$. For the  observers in $\mathcal{O}^\prime$, we calculate the amplitude for an antiparticle to propagate  from $y$ to $x$. Causality requires that these two amplitudes differ at most by a phase~\cite{Ahluwalia2014}
\begin{align}
 \mbox{Amp}(x\to y,~\mbox{particle})&\vert_{\mathcal{O}} 
  \nonumber\\
 = e^{i \theta}
\mbox{Amp}(y\to x, &~\mbox{antiparticle})\vert_{\mathcal{O^\prime}}
\label{eq:Amp} 
\end{align}
where $\theta\in \Re$.\footnote{The `amplitudes' in (\ref{eq:Amp}) are dimensionful. It is an artefact of the fact that creation and destruction operators in the quantum field theory -- for historical reasons -- carry dimensions.} An explicit calculation of the amplitudes in (\ref{eq:Amp}) using spin sums (\ref{eq:12}) and (\ref{eq:34}) while keeping track of the space-like nature of the $(x-y)$ separation, yields $e^{i\theta}=1$. Thus, we must select the bosonic statistics. This is an unavoidable consequence of introducing the new dual and adjoint in the 
$\mathfrak{a}(x)$-$\gdualn{\mathfrak{a}}(x)$ system. It evades the
spin-statistics theorem~\cite{Streater:1989vi,Duck:1997ua}. The crevice in the theorem is the following: the new adjoint renders the interacting Hamiltonian non-hermitian in a manner that still keeps the spectrum real.

With the  $\mathfrak{a}(x)$-$\gdualn{\mathfrak{a}}(x)$  system now fully specified,   the Feynman-Dyson propagator for the new bosons, up to a factor of $i$, is simply the vacuum expectation value of the time order product
\begin{align}
\langle~\vert \mathfrak{T}[ \mathfrak{a}(x^\prime) \gdualn{\mathfrak{a}}(x)]\vert~\rangle=&\langle~\vert\mathfrak{a}(x^\prime)\gdualn{\mathfrak{a}}(x)\vert~\rangle
\theta(t^\prime-t)\nonumber\\
&+\langle~\vert \gdualn{\mathfrak{a}}(x)\mathfrak{a}(x^\prime)\vert~\rangle
\theta(t-t^\prime)\nonumber
\end{align}
with $\vert~\rangle$ as the vacuum state and $\mathfrak{T}$ as the time ordering operator. The two vacuum expectation values that appear in the right-hand side of the above expression evaluate, respectively to
\begin{align}
\langle~\vert\a(x')\da(x)\vert~\rangle=\int
\frac{d^3 p}{(2\pi)^3} 
&\left(\frac{1}{2 E}\right)e^{-i p\cdot (x^\prime-x)}\nonumber\\
&\times(\gamma_\mu p^\mu+m\I)
\end{align}
and
\begin{align}
\langle~\vert\da(x)\a(x')\vert~\rangle=\int
\frac{d^3 p}{(2\pi)^3} 
&\left(
\frac{1}{2 E} \right)e^{-i p\cdot (x-x^\prime)}\nonumber\\
&\times \left[ -(\gamma_\mu p^\mu-m\I)\right]
\end{align}
where we have used the spin sums (\ref{eq:12}) and (\ref{eq:34}). Substituting the two vacuum expectation values into the expression for 
$\langle~\vert \mathfrak{T}[ \mathfrak{a}(x^\prime) \gdualn{\mathfrak{a}}(x)]\vert~\rangle$  and using the integral representation of the step function
\begin{equation}
\theta(t)=\frac{1}{2\pi i}\int^{\infty}_{-\infty} ds \frac{e^{ist}}{s-i\epsilon},
\end{equation}
we obtain 
\begin{align}
\langle~\vert \mathfrak{T}[ \mathfrak{a}(x^\prime) \gdualn{\mathfrak{a}}(x)]\vert~\rangle=&\frac{i}{(2\pi)^{4}}\int d^{4}q\,
e^{-iq\cdot(x^\prime-x)}\nonumber\\
&\times\frac{\gamma_{\mu}q^{\mu}+m\I}{q\cdot q-m^{2}+i\epsilon}.
\end{align}
Consequently, the mass dimension of the new field 
-- as referred to the $\mathfrak{a}(x)$-$\gdualn{\mathfrak{a}}(x)$ 
system --
is three half,
with the Lagrangian density 
\begin{equation}
\mathcal{L}(x)=\da(x)\left(i\gamma^{\mu}\partial_{\mu}-m\I\right)\a(x)
\end{equation}
from which we obtain the canonical equal-time commutators (thus establishing locality)
\begin{align}
&\left[\a(t,\x),\a(t,\y)\right]=\O =  \left[\mathfrak{p}(t,\x),\mathfrak{p}(t,\y)\right], \\
&\left[\a(t,\x),\mathfrak{p}(t,\y)\right]=i\delta^{3}(\x-\y)\I,
\end{align}
where $\mathfrak{p}(x)=i\da(x)\gamma_{0}$ is the conjugate momentum. The free Hamiltonian evaluates to
\begin{align}
H_{0}=&\int d^{3}p\sqrt{|\p|^{2}+m^{2}}\nonumber\\
&\times\left[\sum_{i=1,2}a^{\dag}_{i}(\p)a_{i}(\p)+\sum_{i=3,4}b_{i}(\p)b^{\dag}_{i}(\p)\right]. \nonumber 
\end{align}
The free Hamiltonian is positive-definite if and only if the annihilation and creation operators satisfy bosonic statistics. 

The above expression can be rewritten as
\begin{align}
H_{0}=& 2\delta^3(0) \int d^3p\, \sqrt{|\p|^2 +m^2} 
 + \int d^{3}p\sqrt{|\p|^{2}+m^{2}}\nonumber\\
&\times\left[\sum_{i=1,2}a^{\dag}_{i}(\p)a_{i}(\p)+\sum_{i=3,4}b^\dagger_{i}(\p)b_{i}(\p)\right].\nonumber
\end{align}
Since 
\begin{equation}
\delta^3(\p) = \frac{1}{(2\pi)^3} \int d^3x \exp(i\p\cdot\x),\nonumber
\end{equation}
$\delta^3(0)$ may be replaced by $\left[1/(2 \pi)^3\right]\int d^3x$. With this observation, the zero-point energy takes the form
\begin{equation}
H_\text{vac} = 4 \times \frac{1}{(2\pi)^3}\int d^3x \int d^3 p \,\frac{1}{2} 
\sqrt{|\p|^{2}+m^{2}}.
\end{equation}
In natural units $\hbar =1$ implies $2\pi=h$, so $H_{\text{vac}}$ represents a zero-point energy assignment of $+\frac{1}{2}\sqrt{|\p|^2+m^2}$ to each unit-size cell 
$(1/h^3) d^3x \,d^3p$ in the sense of statistical mechanics. The factor of $4$ is consistent with  the four degrees of freedom (when spinorial degrees of freedom associated with particles and antiparticles are accounted for).
The second term in $H$  shows that each of the four degrees of freedom contributes exactly the same energy to the field for a given momentum $\p$. If for each of the SM Dirac fermions there exists a new boson of equal mass then the zero-point energies exactly cancel. This is a natural consequence of our formalism.

\section{Unitary evolution and scattering theory}

The bilinear invariants constructed from $\mathfrak{a}(x)$ and 
$\gdualn{\mathfrak{a}}(x)$ that enter the interacting $H$ are not hermitian; see (\ref{eq:25}) and (\ref{eq:26}) below. Despite this, the interacting $H$ still carries a real spectrum provided the norm of states is appropriately modified.

To develop the formalism, we introduce 
\begin{equation}
{H}^{\#}\stackrel{\textrm{def}}{=}  \eta^{-1}{H}^{\dag}\eta={H} \label{eq:gen_Herm_H}
\end{equation}
where $\eta$ is a hermitian operator, to be determined below. The work of Mostafazadeh then 
assures that ${H}$ has a real spectrum provided that 
its eigenstates $|\psi\rangle$ have non-vanishing $\eta$-norms $\langle\psi|\eta|\psi\rangle\neq0$. To see this, we note that (\ref{eq:gen_Herm_H}) yields
\begin{align}
0&=\langle\psi|(H^{\dag}\eta-\eta H)|\psi\rangle\nonumber\\
&=(E^{*}-E)\langle\psi|\eta|\psi\rangle.
\end{align}
Therefore, when $\langle\psi|\eta|\psi\rangle\neq0$, the eigenvalues are real~\cite{Mostafazadeh:2001jk,LeClair:2007iy,Bender:1998ke,RevModPhys.15.175}. }

To preserve unitarity, we construct $\eta$ to have the following properties~\cite{LeClair:2007iy}
\begin{equation}
\eta a_{i}(\p)\eta^{-1} =  a_{i}(\p),\quad \eta b_{i}(\p)\eta^{-1}= -b_{i}(\p).\label{eq:eta}
\end{equation} 
That is, $\eta$ commutes with $a_i(\p)$, and anti-commutes with $b_i(\p)$.
Combined with the definition of the dual (\ref{eq:nd}), this gives a relation between $\gdualn{\mathfrak{a}}(x)$ and  $\overline{\mathfrak{a}}(x)$
 \begin{equation}\eta\gdualn{\mathfrak{a}}(x)\eta^{-1}=\overline{\mathfrak{a}}(x) . \label{eq:25}
 \end{equation}
Similarly, the bilinear invariant $\gdualn{\mathfrak{a}}(x)\mathfrak{a}(x)$ is related to its hermitian conjugate by
\begin{equation}
\eta[\gdualn{\mathfrak{a}}(x)\mathfrak{a}(x)]\eta^{-1}=[\gdualn{\mathfrak{a}}(x)\mathfrak{a}(x)]^{\dag}.\label{eq:26}
\end{equation}
{\textcolor{black}{Since $[\eta,H_{0}]=\O$, the free Hamiltonian is hermitian. But in general, the full interacting Hamiltonian is non-hermitian and does not commute with $\eta$, so we require~(\ref{eq:gen_Herm_H}) to ensure the reality of the spectra.}} The simplest contributions from spin-half bosons to $H$ are of the form $\da(x)\a(x)$ and $\da(x)\gamma^{5}\a(x)$.

An explicit solution for $\eta$ that satisfies the above requirements may be constructed following~\cite{Robinson:2009xm}. To do that, we introduce
\begin{equation}
\eta \stackrel{\mathrm{def}}{=} e^{i \alpha \chi},\quad\alpha\in \R,
\end{equation}
where
\begin{equation}
\chi=\int d^{3}p\sum_{i=3,4}b^{\dag}_{i}(\p)b_{i}(\p).
\end{equation}
Thus, $\eta$ is a state-dependent relative phase  between the 
particle-antiparticle sector.
A short calculation reveals that
\begin{equation}
\eta b^{\dagger}_{i}(\p,\sigma)\eta^{-1} = e^{i \alpha} b^{\dagger}_{i}(\p,\sigma)
\end{equation}
and helps us fix $\alpha$.
For equation (\ref{eq:eta}) to be satisfied, we must have $\alpha= \pi$ (modulo $\pm$ integral multiples of $2\pi$). This gives us $\eta=e^{i \pi \chi}$, which  is hermitian and unitary.
To see that $\eta$ is indeed hermitian, notice $\eta^\dagger = e^{-i \pi \chi^\dagger}$, but for bosons $\chi^\dagger = \chi$; therefore $\eta^\dagger = e^{- i \pi \chi}$.  Now $\chi$ acting on physical states $\vert\psi\rangle$ gives $n$ where $n$ is either zero, or an integer: $\chi \vert \psi \rangle = n \vert \psi \rangle $. Therefore,  $\eta^\dagger \vert \psi\rangle = e^{- i \pi n} \vert \psi\rangle$, while $\eta \vert \psi\rangle = e^{ i \pi n} \vert \psi\rangle$. Thus while acting on physical states $\eta^\dagger = \eta$, since $e^{-i\pi}$ = $e^{i\pi}$. It is readily seen that $\eta^2 \vert \psi \rangle = \vert \psi \rangle$, so $\eta^2 = \I$. Consequently, while acting on $\vert \psi \rangle$: $\eta^\dagger=\eta^{-1}=\eta$.

{\textcolor{black}{We now develop the scattering theory for spin-half bosons.}}
{\textcolor{black}{Since for the states with time evolution governed by the generalized hermitian Hamiltonian \textendash~ of the class defined above in equation~ (\ref{eq:gen_Herm_H})   \textendash~ their hermitian inner-product is not invariant under time-translation
\begin{equation}
\langle \phi\vert\psi\rangle\big\vert_{t>0}=\langle\phi\vert e^{iH^{\dag}t}e^{-iHt}|\psi\rangle\neq\langle \phi|\psi\rangle\big\vert_{t=0}.
\end{equation} 
It is necessary to introduce the $\eta$-norm~\cite{Mostafazadeh:2001jk,LeClair:2007iy}
\begin{equation}
\langle\phi|\psi\rangle_{\eta} \stackrel{\mathrm{def}}{=} \langle\phi|\eta|\psi\rangle\label{eq:c_prod}
\end{equation}
{\textcolor{black}{where $\eta=e^{i\pi\chi}$.}}  At $t>0$, we find
\begin{equation}
i\frac{d}{dt}\langle\phi|\eta|\psi\rangle\big\vert_{t>0}=\langle\phi|(\eta H-H^{\dag}\eta)|\psi\rangle\big\vert_{t>0}.
\end{equation}
Therefore, the $\eta$-norm is invariant if and only if $H$ satisfies~(\ref{eq:gen_Herm_H}).}}

Using the $\eta$-norm, the $S$-matrix is given by
\begin{eqnarray}
S_{\beta\alpha}&\equiv&\langle\beta_{\text{out}}|\alpha_{\text{in}}\rangle_{\eta}\nonumber\\
&=&\lim_{\tau\rightarrow\infty}\lim_{\tau_{0}\rightarrow-\infty}\langle\beta_{0}|\Omega^{\dag}(\tau)\eta\Omega(\tau_{0})|\alpha_{0}\rangle
\label{eq:smatrix}
\end{eqnarray}
where $\Omega(\tau)=e^{iH\tau}e^{-iH_{0}\tau}$ \textcolor{black}{and we have taken the free Hamiltonian to be hermitian. }Therefore,
\begin{align}
S_{\beta\alpha}=\lim_{\tau\rightarrow\infty}\lim_{\tau_{0}\rightarrow-\infty}\langle\beta_{0}|U(\tau,\tau_{0})|\alpha_{0}\rangle
\end{align}
\textcolor{black}{where}
\begin{align}
U(\tau,\tau_{0})=\eta\, e^{iH_{0}\tau}e^{-iH(\tau-\tau_{0})}e^{-iH_{0}\tau_{0}}. \label{eq:U_tau}
\end{align}
In obtaining~(\ref{eq:U_tau}), we have used the identity 
\begin{equation}
e^{iH^{\dag}\tau}\eta e^{-iH\tau}=\eta \label{eq:eta_Ht}
\end{equation} which is equivalent to~(\ref{eq:gen_Herm_H}).

Taking $H=H_{0}+V$ and differentiating $U(\tau,\tau_{0})$ with respect to $\tau$, we obtain
\begin{align}
&i\frac{d}{d\tau}U(\tau,\tau_{0})=\left[\eta V(\tau)\eta^{-1}\right]U(\tau,\tau_{0}), \label{eq:dU1}\\
& V(\tau)=e^{iH_{0}\tau}Ve^{-iH_{0}\tau}.\label{eq:dU2}
\end{align}
{\textcolor{black}{Using the initial condition $U(\tau_{0},\tau_{0})=\eta$, the solution to $U$ is given by
\begin{align}
U(\tau,\tau_{0})&=\eta-i\int^{\tau}_{\tau_{0}}d\tau'\left[\eta V(\tau')\eta^{-1}\right]U(\tau',\tau_{0})\nonumber\\
&=\eta\left[I-i\int^{\tau}_{\tau_{0}}d\tau' V(\tau')+\cdots\right].
\end{align}
Taking $V(t)=\int d^{3}x\mathcal{H}(x)$, the Dyson series is
\begin{align}
S_{\beta\alpha}=&\eta_{\beta\alpha}+\sum^{\infty}_{n=1}\frac{(-i)^{n}}{n!}\int d^{4}x_{1}\cdots d^{4}x_{n}\nonumber\\
&\times\langle\beta_{0}|\eta \mathfrak{T}\left[\mathcal{H}(x_{1})\cdots\mathcal{H}(x_{n})\right]|\alpha_{0}\rangle
\end{align}
where $\eta_{\beta\alpha}\equiv\langle\beta_{0}|\eta|\alpha_{0}\rangle$.}}

Next, we address the issue of unitarity. For hermitian Hamiltonians $H^{\dag}=H$, unitarity implies $\int d\beta S^{\dag}_{\gamma\beta}S_{\beta\alpha}=\delta(\gamma-\alpha)$ where $\int d\beta$ integrates over momentum, sums over particle species and internal degrees of freedom. As for generalized hermitian Hamiltonians, we have $H^{\#}=H$ so it suggests the following generalization to unitarity~\cite{Simon:2018zrj}
\begin{equation}
\int d\beta S^{\#}_{\gamma\beta}S_{\beta\alpha}=\delta(\gamma-\alpha),\label{eq:unit}
\end{equation}
where the matrix element of $S^{\#}_{\gamma\beta}$ is given by
\begin{align}
&S^{\#}_{\gamma\beta}=\lim_{\tau\rightarrow\infty}\lim_{\tau_{0}\rightarrow-\infty}\langle\gamma_{0}|U^{\#}(\tau,\tau_{0})|\beta_{0}\rangle, \\
& U^{\#}(\tau,\tau_{0})\equiv \eta^{-1}U^{\dag}(\tau,\tau_{0})\eta.
\end{align}
To prove~(\ref{eq:unit}), 
we start with orthonormality and completeness relations
associated with the free (and hermitian) Hamiltonian $H_0$, 
 and then implement evolution due to the full Hamiltonian $H$ satisfying (\ref{eq:gen_Herm_H}) as  follows:
\begin{equation}
\int d\alpha |\alpha_{0}\rangle\langle\alpha_{0}|=I,\quad \langle\beta_{0}|\alpha_{0}\rangle=\delta(\beta-\alpha).\label{eq:completeness}
\end{equation}
{\textcolor{black}{
Therefore, equation~(\ref{eq:unit}) is satisfied when 
\begin{equation}
U^{\#}(\tau,\tau_{0})U(\tau,\tau_{0})=I \label{eq:U_hash_U}
\end{equation}
for all $\tau$ and $\tau_{0}$. Rewriting $U(\tau,\tau_{0})$ as
\begin{equation}
U(\tau,\tau_{0})=e^{iH_{0}\tau}e^{-iH^{\dag}(\tau-\tau_{0})}e^{-iH_{0}\tau_{0}}\eta,
\end{equation}
we find
\begin{align}
&U^{\#}(\tau,\tau_{0})U(\tau,\tau_{0})\nonumber\\
&=e^{iH_{0}\tau_{0}}\left[e^{iH(\tau-\tau_{0})}\eta e^{-iH^{\dag}(\tau-\tau_{0})}\right]e^{-iH_{0}\tau_{0}}\eta.
\end{align}
Taking the inverse of~(\ref{eq:eta_Ht}), we get
\begin{equation}
e^{iH(\tau-\tau_{0})}\eta^{-1}e^{-iH^{\dag}(\tau-\tau_{0})}=\eta^{-1}.
\end{equation}
Since $\eta=\eta^{-1}$, we obtain~(\ref{eq:U_hash_U}) as required.
}}

{\textcolor{black}{From~(\ref{eq:dU1}) and~(\ref{eq:dU2}), we obtain
\begin{equation}
-i\frac{d}{d\tau}U^{\#}(\tau,\tau_{0})=U^{\#}(\tau,\tau_{0})V^{\dag}(\tau).
\end{equation}
Following the same procedure in deriving the Dyson series for $S_{\beta\alpha}$, we obtain
\begin{align}
S^{\#}_{\gamma\beta}=&\eta_{\gamma\beta}+\sum^{\infty}_{n=1}\frac{i^{n}}{n!}\int d^{4}x_{1}\cdots d^{4}x_{n}\nonumber\\
&\times\langle\gamma_{0}|\eta\mathfrak{T}\left[\mathcal{H}^{\dag}(x_{n})\cdots \mathcal{H}^{\dag}(x_{1})\right]|\beta_{0}\rangle.\nonumber\label{eq:dyson2}
\end{align}
}}

{\textcolor{black}{
Normalizing the $S$-matrix as 
\begin{align}
&S_{\beta\alpha}\equiv \eta_{\beta\alpha}-2\pi iM_{\beta\alpha}\delta^{4}(p_{\beta}-p_{\alpha}),\\
&S^{\#}_{\gamma\beta}\equiv \eta_{\gamma\beta}+2\pi iM^{\#}_{\gamma\beta}\delta^{4}(p_{\gamma}-p_{\beta}),
\end{align}
equation~(\ref{eq:unit}) becomes
\begin{align}
&i\int d\beta\left[\eta_{\gamma\beta}M_{\beta\alpha}\delta^{4}(p_{\beta}-p_{\alpha})-M^{\#}_{\gamma\beta}\eta_{\beta\alpha}\delta^{4}(p_{\gamma}-p_{\beta})\right]\nonumber\\
&=2\pi\int d\beta
\left[\delta^{4}(p_{\beta}-p_{\gamma})\delta^{4}(p_{\beta}-p_{\alpha})M^{\#}_{\gamma\beta}M_{\beta\alpha}\right]. \label{eq:im}
\end{align}
Setting $\gamma=\alpha$, the term $\delta^{4}(p_{\beta}-p_{\alpha})$ cancels from both side of~(\ref{eq:im}) so we obtain
\begin{align}
&i\int d\beta\left(\eta_{\alpha\beta}M_{\beta\alpha}-M^{\#}_{\alpha\beta}\eta_{\beta\alpha}\right)\nonumber\\
&=2\pi\int d\beta\left[\delta^{4}(p_{\beta}-p_{\alpha})
M^{\#}_{\alpha\beta}M_{\beta\alpha}\right].\label{eq:mm}
\end{align}
Equations~(\ref{eq:im}) and~(\ref{eq:mm}) can be seen as a generalization to the optical theorem. In the subsequent section, we will demonstrate that~(\ref{eq:mm}) is satisfied for a model of Yukawa interaction involving spin-half bosons.}}

To compute physical observables, we first need to define the transition probability for the process $\alpha\rightarrow\beta$. Since the unitarity of the $S$-matrix is defined using the generalized hermitian adjoint $\#$, the transition probability should be appropriately modified. In a finite volume $V$ with duration $T$, we define the transition probability to be~\cite{Weinberg:1995mt}

\begin{align}
\textcolor{cyan}{P(\alpha\rightarrow\beta)\equiv}&\textcolor{cyan}{\left[\frac{(2\pi)^{3}}{V}\right]^{N_{\alpha}+N_{\beta}}\frac{VT}{(2\pi)^{2}}\left(\wp_{\beta\alpha}M^{\#}_{\alpha\beta}M_{\beta\alpha}\right)}\nonumber\\
&{\textcolor{cyan}{\times\left[\delta^{3}_{V}(p_{\beta}-p_{\alpha})\delta_{T}(E_{\beta}-E_{\alpha})\right].}} \label{eq:prob}
\end{align}
Because $M^{\#}_{\alpha\beta}M_{\beta\alpha}$ is not guaranteed to be positive-definite, we multiply it by a constant phase $\wp_{\beta\alpha}\in\Re$ to ensure that $P(\alpha\rightarrow\beta)\geq0$. {From the definition of $\eta$ and the Dyson series of $S_{\beta\alpha}$ and $S^{\#}_{\alpha\beta}$, we find that the \textcolor{cyan}{$M^{\#}_{\alpha\beta}=e^{i\pi(n_{\beta}-n_{\alpha})}M^{\dag}_{\alpha\beta}$} where $n_{\alpha}$ and $n_{\beta}$ are the number of spin-half anti-bosons present in states $|\alpha_{0}\rangle$ and $|\beta_{0}\rangle$ respectively. Therefore, by choosing \textcolor{cyan}{$\wp_{\alpha\beta}=e^{-i\pi(n_{\beta}-n_{\alpha})}$}, equation~(\ref{eq:prob}) reduces to the standard transition probability and the physical observables defined using hermitian conjugation can be applied without difficulty.

\subsection{Phenomenologies}

We now consider two spin-half bosonic and fermionic fields of equal mass
coupled to a real scalar field $\phi(x)$
\begin{align}
V(t)=g\int d^{3}x\left[\overline{\psi}(x)\psi(x)+\da(x)\a(x)\right]\phi(x)\label{eq:Yukawa}
\end{align}
where $\psi(x)$ is a fermionic field whose expansion coefficients are given by~(\ref{eq:xi12}) and~(\ref{eq:xi34}). 
{\textcolor{black}{By the construction of $\eta$, it commutes with the fermionic and scalar fields
\begin{equation}
\left[\eta,\psi(x)\right]=\left[\eta,\phi(x)\right]=O.
\end{equation}
Therefore, the Yukawa interaction~(\ref{eq:Yukawa}) is a generalized hermitian operator $\eta^{-1}V^{\dag}(t)\eta=V(t)$.
}}

We now demonstrate that the Yukawa interaction satisfies~(\ref{eq:unit}), the generalized unitarity relation. Taking the initial state to be a single scalar boson, {\textcolor{black}{the matrix element of $\eta$ is $\eta_{\phi\beta}=\eta_{\beta\phi}=\delta(\beta-\phi)$ so~(\ref{eq:mm}) becomes}}
\begin{equation}
i(M_{\phi\phi}-M^{\#}_{\phi\phi})=2\pi\int d\beta\left[\delta^{4}(p_{\beta}-p_{\phi})M^{\#}_{\phi\beta}M_{\beta\phi}\right].\label{eq:Im}
\end{equation}
We will evaluate both sides of~(\ref{eq:Im}) independently. Starting with the left-hand side, the matrix elements $M_{\phi\phi}$ and $M^{\#}_{\phi\phi}$ are obtained by evaluating the amplitude for $\phi\rightarrow\phi'$ and subsequently setting $\phi'=\phi$. 

Using the solution of $\eta$, we obtain $\eta|\phi\rangle=|\phi\rangle$ where $|\phi\rangle$ denotes a single scalar boson state. Setting $\phi=\phi'$, we find $M^{\dag}_{\phi\phi}=M^{\#}_{\phi\phi}$. Therefore, we may rewrite~(\ref{eq:Im}) as
\begin{equation}
\text{Im}(M_{\phi\phi})=-\pi\int d\beta\left[\delta^{4}(p_{\beta}-p_{\phi})M^{\#}_{\phi\beta}M_{\beta\phi}\right].
\end{equation}
Here, $\mbox{Im}(M_{\phi\phi})$ is proportional to the imaginary part of the loop corrections to the scalar propagator~\cite{Schwartz:2014sze}. Due to the difference in statistics and the fact that the spin-half bosons and fermions have equal mass, the bosonic and fermionic loop corrections identically cancel to all orders in perturbation so $\text{Im}(M_{\phi\phi})=0$. Next, we evaluate the right-hand side of~(\ref{eq:Im}). It expands to
\begin{align}
&\int d\beta\left[\delta^{4}(p_{\beta}-p_{\alpha})M^{\#}_{\phi\beta}M_{\beta\phi}\right]\nonumber\\
&=\int d\beta_{\psi}
\sum_{\text{spins}}M^{\#}_{\phi(\overline{\psi}\psi)}M_{(\overline{\psi}\psi)\phi}\delta^{4}(p_{\,\overline{\psi}}+p_{\psi}-p_{\phi}) \nonumber\\
&+\int d\beta_{\mathfrak{a}}\sum_{\text{spins}}
M^{\#}_{\phi(\gdualn{\mathfrak{a}}\mathfrak{a})}M_{(\gdualn{\mathfrak{a}}\mathfrak{a})\phi}\delta^{4}(p_{\gdualn{\mathfrak{a}}}+p_{\mathfrak{a}}-p_{\phi})\label{eq:db}
\end{align}
where $d\beta_{\psi}= d^{3}p_{\psi}d^{3}p_{\,\overline{\psi}}$ and $d\beta_{\mathfrak{a}}= d^{3}p_{\mathfrak{a}}d^{3}p_{\gdualn{\mathfrak{a}}}$. The first and second term of~(\ref{eq:db}) comes from the fermionic and bosonic interactions respectively. 
{\textcolor{black}{
Acting $\eta$ on the relevant states in~(\ref{eq:db}), we obtain
\begin{align}
\eta|\phi\rangle=|\phi\rangle,\quad \eta|\overline{\psi}\psi\rangle=|\overline{\psi}\psi\rangle,\quad \eta|\da\a\rangle=-|\da\a\rangle. \nonumber
\end{align}
where $|\overline{\psi}\psi\rangle$ and $|\da\a\rangle$ denote the pair of spin-half fermion-anti-fermion and boson-anti-boson states respectively. The actions of $\eta$ on $|\overline{\psi}\psi\rangle$ and $|\da\a\rangle$ have opposite signs which induces a relative phase in~(\ref{eq:db}). In effect, we find
{\textcolor{black}{
\begin{equation}
M^{\#}_{\phi(\overline{\psi}\psi)}=M^{\dag}_{\phi(\overline{\psi}\psi)},\quad
M^{\#}_{\phi(\gdualn{\mathfrak{a}}\mathfrak{a})}=-M^{\dag}_{\phi(\gdualn{\mathfrak{a}}\mathfrak{a})}.
\end{equation}}}
Using the decay rate formula~\cite{Weinberg:1995mt}
\begin{align}
\Gamma(1\rightarrow1'2')=2\pi\int & d\beta_{1',2'}\sum_{\text{spins}}|M_{(1'2')1}|^{2}\nonumber\\
&\times\delta^{4}(p_{1'}+p_{2'}-p_{1}),\label{eq:decay}
\end{align}
where $d\beta_{1',2'}=d^{3}p'_{1}d^{3}p'_{2}$, we obtain
\begin{align}
&\int d\beta\left[\delta^{4}(p_{\beta}-p_{\alpha})M^{\#}_{\phi\beta}M_{\beta\phi}\right]\nonumber\\
&\qquad\qquad=\frac{1}{2\pi}\left[\Gamma(\phi\rightarrow\overline{\psi}\psi)-\Gamma(\phi\rightarrow\gdualn{\mathfrak{a}}\mathfrak{a})\right]
\end{align}
so the unitarity condition becomes
\begin{equation}
\text{Im}(M_{\phi\phi})=\frac{1}{2}\left[\Gamma(\phi\rightarrow\gdualn{\mathfrak{a}}\mathfrak{a})-\Gamma(\phi\rightarrow\overline{\psi}\psi)\right].\label{eq:pseudo_u}
\end{equation}
Since the bosons and fermions have the same mass and their quantum fields are expanded the same spinors, we have $\Gamma(\phi\rightarrow\overline{\psi}\psi)=\Gamma(\phi\rightarrow\gdualn{\mathfrak{a}}\mathfrak{a})$. Therefore, the right-hand side of~(\ref{eq:pseudo_u}) identically vanishes  to all orders in perturbation. This is in agreement with $\text{Im}(M_{\phi\phi})=0$ so the theory is unitary.}}

\section{Conclusion} Spin-statistics theorem has a very venerated position in the theory of quantum fields, as does Dirac equation. The works in the past have shown that the Dirac formalism with mass dimension three half, needs to be updated with a new formalism of mass dimension one fermions. Here we have argued that the spin-statistics theorem can also be evaded by exploiting the freedom in defining new dual for spinors and the ensuing adjoint. This results in a new quantum theory of spin half bosons. We then formulated a generalisation of the S-matrix to calculate some elementary processes.

\acknowledgments
We are grateful to James Brister, Julio M. Hoff~da~Silva, Ali Mostafazadeh, Takaaki Nomura, Zheng~Sun and Cong Zhang for useful discussions. CYL is supported by The Sichuan University Post-doctoral Research Fund No.~2022SCU12119.

\providecommand{\href}[2]{#2}\begingroup\raggedright\endgroup

\end{document}